\documentclass[11pt,a4paper]{article}                           
\usepackage[bookmarksopen, bookmarksnumbered, bookmarksopenlevel=2]{hyperref}
\usepackage{tikz}
\usetikzlibrary{cd}
\usepackage{tikz-3dplot}
\usetikzlibrary{calc}
\usetikzlibrary{decorations} %
\usepackage[UKenglish]{babel}
\usepackage{amsmath}
\usepackage{amssymb}
\usepackage{mathtools}
\usepackage{graphicx}
\usepackage{wrapfig}
\usepackage{hhline}
\usepackage[bf]{caption}
\usepackage{cite}
\usepackage[vcentermath]{youngtab}
\usepackage{enumerate}

\DeclareFontFamily{OT1}{rsfs10}{}
\DeclareFontShape{OT1}{rsfs10}{m}{n}{ <-> rsfs10 }{}
\DeclareMathAlphabet{\mathscript}{OT1}{rsfs10}{m}{n}

\numberwithin{equation}{section}

\newcommand{\bbR}{{\mathbb{R}}}
\newcommand{\bbP}{{\mathbb{P}}}
\newcommand{\bbZ}{{\mathbb{Z}}}

\newcommand{\KB}{{\overline{K}_B}}

\def\o{\omega}

\def\cC{{\mathcal C}}
\def\G{\Gamma}

\def\S{\Sigma}

\def\cE{{\mathcal E}}
\def\s{\sigma}
\def\cO{{\mathcal O}}
\def\cN{{\mathcal N}}
\def\cP{{\mathcal P}}
\def\cM{{\mathcal M}}
\newcommand{\pt}{\partial}

\def\l{\lambda}
\def\cS{{\mathcal S}}

\makeatother
\hypersetup{
    pdftitle={Non-vanishing Heterotic Superpotentials on Elliptic Fibrations},
    pdfauthor={Evgeny I. Buchbinder, Ling Lin, Burt A. Ovrut},
    pdfsubject={Heterotic string, worldsheet instanton, non-perturbative superpotential, vector bundle moduli, Beasley--Witten}
}

\begin{document}

\begin{titlepage}

\renewcommand*{\thefootnote}{\fnsymbol{footnote}}

\title{{\LARGE Non-vanishing Heterotic Superpotentials on Elliptic Fibrations}\\[1em] }
\author{
Evgeny I.~Buchbinder$^{1}$\footnote{evgeny.buchbinder@uwa.edu.au} , \, 
Ling Lin$^{2}$\footnote{lling@physics.upenn.edu} ,\,
Burt A.~Ovrut$^{2}$\footnote{ovrut@elcapitan.hep.upenn.edu}
}

\date{}
\maketitle
\begin{center} { 
${}^1${\it Department of Physics,  \\ The University of Western Australia,\\
35 Stirling Highway, Crawley WA 6009, Australia\\[3mm]
${}^2$Department of Physics and Astronomy, \\
University of Pennsylvania, \\
209 S 33rd Street, Philadelphia, PA 19104-6396, USA 
}}\\
\end{center}


\vskip 1cm

\begin{abstract}
\noindent
We present models of heterotic compactification on Calabi--Yau threefolds and compute the non-perturbative superpotential for vector bundle moduli.
The key feature of these models is that the threefolds, which are elliptically fibered over del Pezzo surfaces, have homology classes with a unique holomorphic, isolated genus-zero curve.
Using the spectral cover construction, we present vector bundles for which we can explicitly calculate the Pfaffians associated with string instantons on these curves. These are shown to be non-zero, thus leading to a non-vanishing superpotential in the 4D effective action.
We discuss, in detail, why such compactifications avoid the Beasley--Witten residue theorem.

\end{abstract}

\thispagestyle{empty}

\end{titlepage}

\tableofcontents

\section{Introduction}

\renewcommand*{\thefootnote}{\arabic{footnote}}
\setcounter{footnote}{0}


It has been demonstrated that heterotic M-theory \cite{Lukas:1998ew,Lukas:1998tt} and $E_{8} \times E_{8}$ heterotic string theory have vacuum states that, at low energy, can give rise to phenomenologically viable models of $N=1$ supersymmetric particle physics. Examples of such vacua include \cite{Braun:2005bw,Braun:2005zv,Braun:2005nv}, \cite{Bouchard:2005ag,Bouchard:2006dn} and \cite{Anderson:2009mh, Anderson:2011ns, Anderson:2011vy, Anderson:2012yf, Buchbinder:2014qda, Buchbinder:2014sya}. Furthermore, these theories can lead to new inflationary scenarios for the early universe, as well as concrete mechanisms for alternative approaches to cosmology, such as 
``bouncing'' cosmologies via heterotic five-branes, see \cite{Battarra:2014tga,Deen:2016zfr,Deen:2017dpm,Cai:2018ljy} for recent results. Hence, it is imperative to move beyond the four-dimensional low energy limits of these theories and to address, and solve, the fundamental questions that arise from their  compactification from higher-dimensions. Perhaps the foremost of these issues involves calculating the effective potential energies of the compactification moduli and, specifically, to demonstrate 1) that their vacua are stable and 2) the exact mechanism for the spontaneous breaking of $N=1$ supersymmetry. Considerable progress has been made in presenting mechanisms for computing the potential energy and stabilizing the K\"ahler and complex structure moduli arising from the compactification on a Calabi--Yau threefold \cite{Anderson:2009sw,Anderson:2010mh,Anderson:2011cza,Anderson:2011ty} . Similarly, stabilizing the moduli of bulk space five-branes in heterotic M-theory vacua has been discussed in \cite{Donagi:1999jp}. However, as we will now outline in detail, it has been much more difficult to calculate the potentials for the moduli associated with the holomorphic vector bundles on the Calabi--Yau threefold. Such bundles are required, amongst other things, to produce realistic low energy particle spectra. 

Let us be more specific. In compactifications of the heterotic string, moduli fields corresponding to deformations of the vector bundle $V$ can only be stabilized through non-perturbative contributions to the superpotential.
These contributions are generated by worldsheet instantons;  that is, strings wrapping holomorphic, isolated, genus-zero (or rational) curves in the internal space $X$ \cite{Dine:1986zy, Dine:1987bq}.
For viable model building it is imperative, therefore, to construct compactification scenarios in which these instanton contributions give rise to a non-vanishing superpotential. The main challenge in this endeavor is the dependence of the instanton path integral on the Calabi--Yau metric.
Despite significant efforts \cite{Distler:1986wm, Distler:1987ee, Aspinwall:1994uj, Silverstein:1995re, Becker:1995kb, Witten:1999eg, Harvey:1999as, Lima:2001jc, Buchbinder:2002ic, Buchbinder:2002pr, Basu:2003bq, Beasley:2003fx, Beasley:2005iu, Braun:2007tp, Braun:2007xh, Braun:2007vy, Aspinwall:2011us, Buchbinder:2016rmw, Buchbinder:2017azb}, it is usually impossible to determine the explicit form of the instanton contributions due to the limited methods available to write down the metric in explicit models.
However, parts of the superpotential can be computed algebraically.
Schematically, the superpotential contribution of strings wrapping an isolated rational curve $\Gamma$ is  \cite{Witten:1999eg}
\begin{align}\label{eq:schematic_superpotential}
	W(\G) = \exp(i \, A_{\mathbb{C}}(\G)) \frac{\text{Pfaff}( {\cal D}_{F} )}{ \sqrt{ \det ({\cal D}_B) }} \, .
\end{align}
The argument of the exponential factor is the ``complexified'' area $A_{\mathbb{C}}$ of the curve, and only depends on the K\"ahler moduli.
In the fraction, the denominator comes from integrating over the bosonic degrees of freedom in the worldsheet path integral and is independent of the vector bundle moduli.
The numerator comes from integrating over the fermions. Specifically, it is the Pfaffian of the Dirac operator coupled to the restriction of the vector bundle $V$ to the curve $\Gamma$.
While the denominator cannot be computed explicitly, it is known \cite{Witten:1999eg} that the full expression vanishes if and only if the Pfaffian vanishes.
Unlike for the bosonic determinant, the Pfaffian can be calculated with algebraic methods developed in \cite{Buchbinder:2002pr, Buchbinder:2002ic}.
Therefore, the full vector bundle moduli dependence of the instanton contribution \eqref{eq:schematic_superpotential} can be determined algebraically up to a non-zero prefactor. However, this unknown prefactor turns out to be essential for determining the ``full'' superpotential; that is, the instanton contributions summed over all isolated rational curves in $X$.

The reason for this is due to a result of Beasley and Witten. Using a ``residue theorem'' \cite{Beasley:2003fx}, these authors have shown that, for vacua satisfying certain explicit properties which we list below, the worldsheet instanton contributions to the superpotential arising from different isolated rational curves with the same complexified area $A_{\mathbb{C}}$,  
that is, curves in the same homology class, must sum to zero.
Because, in general, the instanton contribution to the superpotentials of individual curves can and will be non-zero, the vanishing of their sum over a homology class has to be the result of non-trivial cancellations due to the prefactors. It follows that any heterotic vacuum that satisfies the assumptions of the Beasley--Witten theorem will have a vanishing superpotential for the vector bundle moduli and, hence, be physically unstable. 
Therefore, to obtain an acceptable heterotic compactification, one must show that it violates at least one of the Beasley--Witten assumptions. 

Formulated in the context of this paper, these assumptions can be stated as follows:
\begin{enumerate}[{1)}]
	\item The compactification space $X$ is a complete intersection Calabi--Yau (CICY) threefold within a higher-dimensional toric ambient space ${\cal{A}}$.
	\item $X$ is ``favorable'' \cite{Anderson:2008uw} with respect to the ambient space; that is, the K\"ahler form $\omega_{X}$ on $X$ is the restriction $\omega|_{X}$ of the K\"ahler form $\omega$ on ${\cal{A}}$.
	\item The holomorphic vector bundle $V$ on $X$ is the restriction $W|_{X}$ of the vector bundle $W$ on the ambient space ${\cal{A}}$.
\end{enumerate}
There have already been a number of heterotic vacua presented in which assumptions 1) and 2) have been violated.
As a consequence, one obtained a non-vanishing instanton superpotential for the vector bundle moduli. 
The first such example was constructed in \cite {Braun:2007tp, Braun:2007xh, Braun:2007vy} and \cite{Buchbinder:2016rmw}. 
This result was generalized to a large number of other vacua in \cite{Buchbinder:2017azb} which, due to their unfavorable embedding in their ambient spaces and, hence, their violation of assumption 2), also had non-zero superpotentials.
However, in all of these examples, the vector bundle descended by restriction to the relevant threefold and, therefore, assumption 3) of the Beasley--Witten theorem was never violated.
Thus, to further explore possible scenarios in which there is a viable mechanism for moduli stablization, an obvious question is:
Can a non-vanishing superpotential for vector bundle moduli be generated on a threefold $X$ that has a favorable toric embedding, thus satisfying assumptions 1) and 2) of the Beasley--Witten theorem, but for which assumption 3) is violated?

The most straightforward setting, in which this question can be definitively answered, is on manifolds $X$ that have a single isolated genus-zero curve $\cE$ within its homology class $[\cE]$.
In this case, there is only a single contribution to the superpotential generated by $[\cE]$, which vanishes if and only if the corresponding Pfaffian does.
Thus, the validity of the residue theorem can be explicitly verified/disproved using algebraic methods \cite{Buchbinder:2002ic, Buchbinder:2002pr} to calculate the Pfaffian.
The purpose of this paper is to demonstrate that on such manifolds, we can indeed construct a broad class of vector bundles whose associated Pfaffians are non-zero. Hence, they present a set of toy models for heterotic compactifications with known non-vanishing superpotentials. In fact, a subset of such vacua were already constructed, and their non-vanishing superpotentials calculated, in \cite{Buchbinder:2002pr, Buchbinder:2002ic}. In this paper we extend and clarify those results. Having done this, we will then demonstrate that in this class of vacua, the vector bundle $V$ on $X$ explicitly cannot descend from a vector bundle on the ambient space---thus violating assumption 3) of the Beasley--Witten theorem and explaining why the superpotential does not vanish.

To this end, we will first review in Section \ref{sec:covers_and_fibrations} the construction of suitable vector bundles via spectral covers on elliptically fibered Calabi--Yau threefolds.
We will show that generic elliptic fibrations over del Pezzo surfaces have homology classes with a single isolated, genus-zero curve.
In Section \ref{sec:superpotentials} we then discuss the algebraic method of computing Pfaffians, and demonstrate for toy spectral cover bundles that the superpotential is explicitly non-zero.
It turns out that these models have underlying Calabi--Yau geometries that actually satisfy the Beasley--Witten assumptions.
To resolve the apparent tension with the residue theorem, we argue in Section \ref{sec:beasley_witten} that our toy models violate assumption 3) of the theorem through the bundle construction.
This argument opens up a novel, algebraic perspective on the residue theorem, which we will highlight in Section \ref{sec:summary} together with other future directions.



\section{Spectral covers and elliptic threefolds over del Pezzo surfaces}\label{sec:covers_and_fibrations}

In this section, we first review the construction of stable vector bundles on elliptic fibrations via spectral data.
Then we will present a class of elliptic fibrations with isolated rational curves that are unique in their homology class.


\subsection{Spectral cover vector bundles}\label{sec:review_spectral_cover}


Throughout these notes, we will consider Calabi--Yau manifolds $X \stackrel{\pi}{\rightarrow} B$ that are elliptically fibered over complex surface $B$.
A distinct feature of elliptic fibrations is the existence of a global section $\s: B \to X$, called the zero section, which can be viewed as an embedding of the base $B$ into $X$.

On elliptically fibered Calabi--Yau manifolds, stable vector bundles with $SU(n)$ structure group can be constructed in terms of spectral data~\cite{Donagi:1997dp, Friedman:1997yq, Friedman:1997ih}. 
It consists of the spectral cover ${\cal C} \subset X$ together with the line bundle ${\cal N}$ on ${\cal C}$. 
The cover $\cal C$ is a divisor in $X$ which is of degree $n$ over the base $B$. 
That is, the restriction $\pi_{\cC} = \pi|_{\cal C} : \cC \to B$ of the elliptic fibration is an $n$-sheeted branched cover.
This means that the homology class of $\cC$ in $H_4( X, {\mathbb Z})$ must be of the form
\begin{align}
[\cC ] = n \s + \pi^{-1} \eta\,, 
\label{2.3}
\end{align}
where $\eta$ is a curve class on the base. We will restrict ourselves to spectral covers satisfying some additional properties. First, we 
require that it is an effective class in $H_4 (X, {\mathbb Z})$. This means that $\cC$ is an actual surface in $X$. 
Equivalently, this means that $\eta$ is an effective curve in $H_2 (B, {\mathbb Z})$. 
Second, we impose that $\cC$ is a smooth irreducible surface. 
These conditions  guarantee that the associated vector bundle is stable. 
Finally, we demand that the spectral cover is positive. 

Let us discuss these conditions more explicitly following \cite{Buchbinder:2002ji, Donagi:2004ia}. 
The set of all effective classes in $H_2 (B, {\mathbb Z})$ is called the Mori cone of $B$.
If the Mori cone is generated 
by the curves $C_k \in H_2 (B, {\mathbb Z})$ then a generic effective curve is of the form 
\begin{align} 
\eta = \sum_k a_k C_k\,, 
\label{2.3.1}
\end{align}
where $a_k$ are non-negative integers.
The irreducibility of the spectral cover can be determined via intersection theory.
For example, if the base is a del Pezzo surface $dP_r$, the spectral cover is irreducible if and only if the following 
conditions are satisfied~\cite{Donagi:2004ia}:
\begin{align}
\begin{split}
	& 1) \, \, \eta \cdot E \geq 0 \, \text{ for any generator $E$ of the Mori cone}
\\
& 2) \, \, \eta - n \, \KB \, \text{ is effective in $B$, with $\KB$ the anti-canonical divisor of  $B$}\, . 
\end{split}
\label{2.3.2}
\end{align}
Finally, the spectral cover is positive if its intersection with every effective curve in $X$ is positive. 

The spectral cover can also be defined by an explicit equation. 
For this we describe $X$ as a hypersurface in a $\bbP^2$-bundle over $B$ given by the Weierstrass equation
\begin{align} 
y^2 z = x^3 + g_2 x z^2 + g_3 z^3\,. 
\label{eq:weierstrass_equation}
\end{align}
For $X$ to be Calabi--Yau, the $\bbP^2$ fiber coordinates $[x:y:z]$ and the coefficients $g_2, g_3$ have to be sections of the following line bundles~\cite{Friedman:1997yq}:
\begin{align}
\begin{split}
	& x \in H^0(B, \cO_B(2\KB)) \, , \qquad y \in H^0(B, \cO_B(3\KB) \, , \qquad z \in H^0(B, \cO_B) \\
	& g_2 \in H^0 (B, \cO_B(4 \KB) )\,, \qquad g_2 \in H^0 (B,  \cO_B(6\KB) )\, .
\end{split}
\label{2.3.4}
\end{align}
In terms of these variables the spectral cover can be represented as the zero set of the equation 
\begin{align} 
f= a_0 {\bf Z}^n +a_2 {\bf X} {\bf Z}^{n-2} +a_3 {\bf Y} {\bf Z}^{n-3} + \dots + \left\{ 
\begin{array}{ll}	
	a_n {\bf X}^{n/2} \, , & \, \text{for even } n\\
	a_n {\bf X}^{n-3} {\bf Y} \, ,& \, \text{for odd } n
\end{array}
\right.
\label{2.3.5}
\end{align}
where $x = {\bf X} {\bf Z}$, $y = {\bf Y}$ and $z = {\bf Z}^3$.
The parameters $a_0, \dots, a_n$ are sections of the following line bundles:
\begin{align}
	a_k \in H^0 (B, \cO (k\,\KB + \eta))\,, \quad k=0, \dots, n\,. 
	\label{2.3.6}
\end{align}
For simplicity we will later consider the case $n=3$ in which the equation for the spectral cover takes the simple form 
\begin{align} 
f = a_0 z + a_2 x + a_3 y\,.
\label{2.3.7}
\end{align}

In addition to the spectral cover it is necessary to specify a line bundle ${\cal N}$ on ${\cal C}$. 
For $SU(n)$ bundles, one can assume that ${\cal N}$ is a restriction of a line bundle on $X$ \cite{Friedman:1997ih, Donagi:1997dp, Buchbinder:2002ji} (which, to simplify notation, we will again denote by ${\cal N}$  ), whose first Chern class\footnote{Here and throughout these notes, we will denote characteristic classes by their Poincar\'{e}-dual homology classes, and cup products by their dual intersection products.} has to be
\begin{align} 
c_1  ({\cal N}) =n \left( \frac{1}{2}+ \l \right) \s + \left(\frac{1}{2} - \l \right) \pi^{-1}(\eta) + \left(\frac{1}{2} + n \l \right)\pi^{-1} (\KB)\,, 
\label{2.4}
\end{align}
where $\lambda$ is, a priori, a rational number.
Since $c_1 ({\cal N}) $ must be an integer class, it follows that either
\begin{align} 
n \  {\rm is} \  {\rm odd}\,,  \qquad  \l =m + \frac{1}{2}
\label{2.5}
\end{align}
or 
\begin{align} 
n \  {\rm is} \  {\rm even}\,,  \qquad  \l =m\,, \quad \eta= \KB \ {\rm mod} \ 2\,,
\label{2.6}
\end{align}
where $m \in {\mathbb Z}$. 

Given a spectral cover $\cC$ and a line bundle $\cN$, 
satisfying the above properties, one can now uniquely construct an $SU (n)$ vector bundle $ V$ on $X$ via the so-called Fourier--Mukai transformation,
\begin{align} 
V= \pi_{1 *} (\pi_2^* \cN \otimes \cP)\,, 
\label{2.7}
\end{align}
where $\pi_1$ and $\pi_2$ are the two projections of the fiber product $X \times_B \cC$ onto the two factors $X$
and $\cC$, and $\cP$ is the Poincar\'{e} line bundle on $X \times_B \cC$. 
To gain some intuition for the Fourier--Mukai transformation, we note that the restriction of $V$ to the elliptic fiber is a sum of $n$ degree zero line bundles ${\cal L}_i$.
On an elliptic curve, such a line bundle is always dual to a divisor of the form $Q_i - p$, where $p$ is the origin, i.e., the point marked by the zero section.
For a spectral cover bundle, the $n$ points $Q_i$ are precisely the intersection of the $n$-sheeted cover $\cal C$ with the elliptic fiber.
For completeness, we collect the Chern classes of such spectral cover bundles:
\begin{align}
\begin{split}
	c_1 (V) & = 0 \,,  \\
	c_2 (V) & =  \pi^{-1} (\eta) \cdot \s -\frac{n^3-n}{24} \left(\pi^{-1} (\KB) \right)^2  +\frac{n}{2} \left( \l^2 -\frac{1}{4} \right)  \pi^{-1}\left( \eta \cdot (\eta - n\,\KB) \right)\,, 
	\\
	c_3 (V) & =  2 \l \, \s \cdot \pi^{-1}\left( \eta \cdot (\eta - n\,\KB) \right) \, .
\end{split}
\label{2.8}
\end{align}
We refer the reader to
the original papers~\cite{Donagi:1997dp, Friedman:1997yq, Friedman:1997ih} for details.

As was discussed in the previous section, the Pfaffian depends on the vector bundle moduli. For the spectral cover bundles under consideration 
the moduli come from two sources: the parameters of the spectral cover $\cC$ and the moduli of the line bundle $\cN$. The general formula 
for the number of moduli of $V$ was derived in~\cite{Buchbinder:2002ji, Buchbinder:2004cs}:
\begin{align} 
n (V ) = (h^0 (X, \cO_X (\cC)) - 1) + h^1(X, \cO_X (\cC))\,, 
\label{2.9}
\end{align}
where the first term gives the parameters of the spectral cover and the second term gives the moduli of the line bundle. 
If the spectral cover is chosen to be positive, it was shown in~\cite{Buchbinder:2002ji} that $h^i (X, \cO_X (\cC)) =0$, $i >0$ 
and the second term in~\eqref{2.9} is zero. Then the bundle moduli come only from the parameters of the spectral cover $a_k$ in eq.~\eqref{2.3.6}. 
Since the spectral cover equation~\eqref{2.3.5} is homogeneous in the parameters it follows that the moduli space of $V$ is the projective space
$\cM (V)= {\mathbb P}^{h^0 (X, \cO_X (\cC)) - 1}$ and the Pfaffian in~\eqref{1.6} is naturally a homogeneous polynomial on $\cM (V)$.


\subsection{Elliptic fibrations with isolated rational curves}


In the rest of the paper we will consider a special class of Calabi--Yau threefolds which are elliptically fibered 
over del Pezzo surfaces $dP_r$, $r=1, \dots, 9$.
They have the feature that they contain a natural set of isolated (or, more precisely, infinitesimally rigid) genus-zero (or rational) curves.
These arise as a set of blow-ups of $\bbP^2$ at $r\leq 8$ points in general position.
For $r=9$, the nine points have to be at the common roots of two cubics in $\bbP^2$.
Each such blow-up introduces an irreducible curve with an independent class $E_i$, $i=1,\dots, r$ in $H_2(dP_r, \bbR)$.
Including the hyperplane class $\ell$ that is inherited from the $\bbP^2$, the dimension of 
$H_2 (dP_r, \bbR)$ is $ r + 1$.
The intersection numbers of these curves are given by 
\begin{align} 
\ell \cdot \ell =1\,, \qquad \ell \cdot E_i =0\,, \qquad E_i \cdot E_j= - \delta_{ij}\,. 
\label{2.1}
\end{align}
The Chern classes of $dP_r$ are
\begin{align} 
c_1 (dP_r)= \overline{K}_{dP_r} =3 \ell -\sum_{i=1}^r E_i\,, \qquad c_2 (dP_r)= 3 +r\, ,
\label{2.2.1}
\end{align}
where the anti-canonical class $\overline{K}_{dP_r} \equiv \overline{K}_r$ satisfies $E_i \cdot \overline{K}_r = 1$, $\overline{K}_r^2 = 9 -r \geq 0$. 
The Mori cone---the set of effective curves---of del Pezzo surfaces with $r \geq 2$ is generated by curves $E$ with
\begin{align}\label{eq:rigid_curves_dP}
	E^2 = -1 \, , \quad E \cdot \overline{K}_r = 1 \, .
\end{align}
For $dP_1$, the generators are $E_1$ and $\ell - E_1$, of which only $E_1$ satisfies \eqref{eq:rigid_curves_dP}.

For any $dP_r$, all curves $E$ satisfying \eqref{eq:rigid_curves_dP} are all isolated genus-zero curves which are unique in their homology class. 
The genus is related to the degree of canonical bundle of $E$, which in turn can be computed via adjunction: $g = 1+\frac{1}{2} \deg\, K_E = 1+\frac{1}{2} E \cdot (E - \overline{K}_r) = 0$.
To see that they are isolated, or more precisely, infinitesimally rigid, note that, as curves embedded inside a surface, the degree of the rank one normal bundle $N_{E/dP}$ is simply given by the self-intersection number of the curve, which is $E^2 =-1$.
Since $E$ has genus 0, i.e., is a $\bbP^1$, this yields $N_{E/dP} = {\cal O}_E(-1)$.
Because its cohomologies are zero, there are no normal deformations of $E$, hence it is infinitesimally rigid.
Finally, let us assume that there exists another curve $E'$ in the same class $[E]$ as $E$.
Since $[E] \cdot [E]=-1$ it follows that $E \cdot E'=-1$. Since both $E$ and $E'$ are effective genus-zero curves it is possible if and only if 
$E= E'$ (two distinct effective curves inside a surface always have a non-negative intersection number unless one of them is reducible and contains the other). 
So $E$ is unique in its homology class in $H_2 (dP_r, {\mathbb R})$. 

Given these isolated rational curves on the base $B = dP_r$, we can now use the zero section $\sigma$ of a \textit{generic} elliptic fibration to obtain such curves in $X$.
The attribute generic is important here, because over del Pezzo surfaces, generic elliptic fibrations are smooth and have no reducible fibers.
In terms of the Weierstrass equation \eqref{eq:weierstrass_equation} of $X$, it means that the coefficients $g_{2/3}$ have to be generic sections of their respective line bundles.

In this setting, the surface $\Sigma = \pi^{-1}( E)$ is smooth as well. 
Clearly, $\Sigma$ is also elliptically fibered by $\pi|_\Sigma : \Sigma \rightarrow E$.
Inside it, we define the curve 
\begin{align} 
\cE =\s \cdot \Sigma\, .
\label{2.11}
\end{align}
Since $\sigma$ defines a copy of the base inside $X$, $\cE$ is simply $E$ inside this copy and obviously has genus 0.
It is easy to see that $\cE$ is isolated in $X$. 
We have already shown above that $E$ is rigid in the base, i.e., the normal bundle in the base direction is $\cO(-1)$.
The zero section $\s$ is also rigid and has no deformations along the fiber direction. This means that 
$\cE$ viewed as a curve in $\Sigma$ has normal bundle $\cO (-1)$. Hence, the normal 
bundle of $\cE$ viewed as a curve in $X$ is an extension of $\cO_{\cE} (-1)$ by $\cO_{\cE} (-1)$. 
The space of possible non-trivial extensions is given by $H^1 (\cE \cong {\mathbb P}^1, \cO) =0$ which means 
that the normal bundle is the trivial extension $\cO_{\cE} (-1) \oplus \cO_{\cE} (-1)$. Hence , $\cE$ infinitesimally rigid inside $X$.

Finally, to see that $\cE$ is unique in its homology class, we first show that the elliptically fibered surface $\Sigma$ is a rational elliptic surface\footnote{In fact, a rational elliptic surface is another name for a $dP_9$ surface. Here, the $dP_9$ is not the base of the elliptic fibration, but appears as a surface inside the elliptic threefold.}, i.e., an elliptically fibered surface that is birational to $\bbP^2$.
Equivalently, it is a smooth elliptic fibration over $\bbP^1$ with generically 12 singular I$_1$ fibers.
For our surface $\Sigma$, the singular fibers are inherited from the threefold $X$ and hence sit at the intersection of $E$ with the discriminant $\Delta \equiv 4\,g_2^3 + 27\,g_3^2 = 0$ of the Weierstrass model \eqref{eq:weierstrass_equation} describing $X$.
For generic choices of $g_{2/3}$ over $B = dP_r$, the generic fiber over $\Delta = 0$ is an I$_1$ fiber.
On the base $B= dP_r$, the curve $\Delta =0$ has class $12\,\overline{K}$.
Using the fact that $E$, as a generator of the Mori cone, intersects $\overline{K}_r$ once, we conclude that $\Sigma$ has precisely 12 singular fibers of type I$_1$.
Since the base of $\Sigma$ is $E \cong \bbP^1$, this proofs that $\Sigma$ is a rational elliptic surface.
By construction \eqref{2.11}, our curve $\cE$ is the restriction of global section of $X$, so it is also a section $\Sigma$.
It is well-known that a section on a rational elliptic surface has self-intersection $-1$.
Hence, 
\begin{align}\label{eq:contradiction_1}
	-1 = \cE \cdot_\Sigma \cE = \cE \cdot_\Sigma (\sigma|_\Sigma) = \cE \cdot_X \sigma \, .
\end{align}

Let us now assume that there is another irreducible curve $\cE'$ in the same homology class as $\cE$.
Then it is clear that the push-forwards onto the base must also be homologous, i.e., $ [\pi(\cE)] = [\pi(\cE')] = [E]$.
Since we have shown above that $[E]$ has the unique representative $E$ on $B$, it follows that $\pi(\cE') = E$.
Therefore, $\cE'$ must be a curve inside $\Sigma = \pi^{-1} (E)$ as well.
Inside $\Sigma$, we can again use the argument that two irreducible distinct curves inside a surface must have non-negative intersection number: $\cE \cdot_\Sigma \cE' \geq 0$.
However, we can again interpret the intersection as one inside $X$:
\begin{align}\label{eq:contradiction_2}
	0 \leq \cE' \cdot_\Sigma \cE = \cE' \cdot_\Sigma (\sigma|_\Sigma) = \cE' \cdot_X \sigma \stackrel{[\cE'] = [\cE]}{=} \cE \cdot_X \sigma \, ,
\end{align}
which is clearly a contradiction to \eqref{eq:contradiction_1}.
Therefore, $\cE$ has to be unique in its homology class.

To conclude, we have shown that any curve $E$ in $dP_r$ with self-intersection number $-1$ and intersecting $\overline{K}_{dP_r}$ at one point is an isolated rational curve.
Lifting it with the zero section into a generic elliptic fibration $X \rightarrow dP_r$, as in~\eqref{2.11}, gives rise to a holomorphic, isolated, genus-zero curve $\cE$ which is 
unique in its homology class.
The existence of such curves makes these manifolds very attractive for constructing heterotic compactifications with non-zero superpotentials, as we will discuss now.


%
%
%


\section{Heterotic compactifications with non-zero superpotentials}\label{sec:superpotentials}

Having established the underlying geometry and bundle construction, we now turn our attention to the computation of superpotentials in heterotic models.
Before discussing explicit examples, we will first review the non-perturbative contribution of worldsheet instantons to the heterotic superpotential.

\subsection{The general structure of the non-perturbative superpotential}\label{sec:general_structure_superpotential}

As extensively studied in a variety of 
papers~\cite{Dine:1986zy, Dine:1987bq, Becker:1995kb, Witten:1999eg, Harvey:1999as, Lima:2001jc},
the effective low-energy field theory of the heterotic string contains a non-perturbative superpotential for moduli fields 
which is generated by worldsheet/open membrane instantons.
The structure of the instantons, as well as the structure of the $N=1$ supermultiplets, 
is slightly different in weakly and strongly coupled heterotic string theories. However, the superpotential has the same general form. 
For concreteness, we will discuss the weakly coupled case where the superpotential is generated by strings wrapping holomorphic curves $\G$ in the compactification space $X$.
As shown in \cite{Dine:1986zy, Dine:1987bq}, the path integrals of these strings vanish due to additional fermionic modes unless $\G$ is an isolated genus-zero curve.
The superpotential is then determined by the
classical Euclidean worldsheet action evaluated on the instanton solution and by the 1-loop determinants of the fluctuations around this solution. 

Concretely, let $\G$ be a holomorphic, isolated, genus-zero curve in $X$. 
Then the general form of the superpotential contribution generated by a string wrapping $\G$ 
is~\cite{ Witten:1999eg, Buchbinder:2016rmw}
\begin{align}
	W (\G)= {\rm exp}\left[ -\frac{A(\G)}{2 \pi \alpha'} + i \int_\G B \right] \times 
	\frac{\text{Pfaff}  ({\bar \partial}_{V_\G (-1)})  }{ [ \det'  ({\bar \partial}_{{\cal O}_\G} )]^2  [ \det  ({\bar \partial}_{{\cal O}_\G(-1) } )]^2 } \times \chi (\G)\,. 
\label{1.1}
\end{align}
The expression in the exponent is the classical Euclidean action evaluated on $\G$. 
The first term $A(\G)$ is the area
of the curve given by 
\begin{align}
A(\G)= \int_\G \omega_X\,, 
\label{1.2}
\end{align}
where $\omega_X$ is the K\"ahler form on $X$. 
The second term contains the heterotic $B$-field which in this expression can be taken 
to be a closed 2-form. 
Then we can expand
\begin{align}
	\omega_X=\sum_{I=1}^{h^{1,1}} t^I \omega_I\,, \qquad B=\sum_{I=1}^{h^{1,1}} \phi^I \omega_I \, ,
\label{1.3}
\end{align}
in a basis  $\omega_I$ of $(1,1)$ forms on $X$, $I=1, \dots, h^{1,1} (X)$.
Defining the complexified K\"ahler moduli $T^I= \phi^I +i \frac{t^I}{2 \pi \alpha'}$, the exponential prefactor becomes
\begin{align}
	e^{i  A_{\mathbb C}(\G)} \equiv e^{i \alpha_I(\G) T^I} \quad \text{with} \quad  \alpha_I(\G) =\int_\G \omega_I\,. 
\label{1.4}
\end{align}
Note that the vectors $\alpha_I(\G)$ depend only on the homology class $[\G] \in H_2(X, \bbR)$ of the curve $\G$, and are linearly independent for curve classes that are linearly independent in $H_2(X, \bbR)$.

The second factor in eq.~\eqref{1.1} is the one-loop contribution which depends on the stable holomorphic vector bundle $V$ on $X$. 
The Pfaffian in the numerator is the square root of the determinant of
the Dirac operator on the curve $\G$, twisted by the vector bundle 
$V_\G (-1) = V|_{\G} \otimes {\cal O}_{\G} (-1)$.
It comes from integrating out the fermions in the worldsheet path integral.
In general, it is a holomorphic function in the bundle moduli of $V$ and the 
complex structure moduli of $X$. 
In the denominator,  $\det'  ({\bar \partial}_{{\cal O}_\G} )$ and  $\det  ({\bar \partial}_{{\cal O}_\G (-1) } )$
come from integrating over bosonic fluctuations and do not depend on $V$.

For completeness we have also included the factor $\chi (\G)$ associated with torsion.
In general, the second homology 
group of $X$ is of the form
\begin{align}
	H_2 (X, {\mathbb Z})= {\mathbb Z}^k \oplus G_{{\rm tor}}\,, \qquad k>0\,, 
\label{1.5}
\end{align}
where ${\mathbb Z}^k $ represents the free part and $G_\text{tor} = \text{Tors}(H_2(X,\bbZ))$ is a finite abelian group which represent the torsion part. 
The origin of the factor $\chi(\G)$ lies in the subtleties surrounding the $B$-field \cite{Witten:1999eg}.
Simply speaking, the field strength $H = dB$ need not to be zero in integer cohomology $H^3(X, \bbZ)$ and can have a non-trivial torsion class.
The value of $[H]$ in $\text{Tors}(H^3)$ corresponds to a discrete choice of the heterotic vacuum.
By Poincar\'{e}-duality and the universal coefficient theorem, $\text{Tors}(H^3) \cong \text{Tors}(H_2) = G_\text{tor} $ on a threefold.
Then the factor $\chi(\G)$ is the ``relative difference'' between $[H]$ and the torsion class of $\G$ arising from a proper treatment of the factor $\exp( \, i \int_\G B)$.
As the (imaginary) exponential of the so-called torsion linking number, $\chi(\G)$ can be interpreted as a character of the abelian group $G_\text{tor}$, which is always non-zero.
Since the models we consider have $G_\text{tor}=0$, all $\chi$'s are equal and we will drop them in the following, although we will comment on their significance in Section \ref{sec:residue_review}.

In general, a given homology class $[\G] \in H_2(X, \bbR)$ contains more than one holomorphic, isolated, genus-zero curve.
The number $n_{[\G]}$ of such curves is computed by the (genus-zero) Gromov-Witten invariant.
All such curves in the same homology class have the same exponential $\exp (i A_{\mathbb{C}} ( [\G] ))$.
However, the 1-loop determinants are in general different.
To find the superpotential contribution, $W([\G])$, associated with the class $[\G]$ we have to sum over all holomorphic, isolated, 
genus-zero curves $\G_j$ in this class. 
\begin{align}
W ([\G])= e^{i A_{\mathbb C} ([ \G] ) }  \sum_{j=1}^{n_{[\G]}} 
\underbrace{ \frac{ \text{Pfaff} ({\bar \partial}_{V_{\G_j} (-1)})  }{ [\det'  ({\bar \partial}_{{\cal O}_{\G_j}} )]^2 [\det ( {\bar \partial}_{{\cal O}_{\G_j} (-1)})]^2} }_{W(\G_j)}\, .
\label{1.6}
\end{align}
Note that the full superpotential $W= \sum_{[\G]} W ([\G])$ is dominated by the contributions from classes with small areas due to the exponential suppression $\exp( i A_{\mathbb C}([\G]) ) \sim \exp(- A([\G]))$.

For heterotic model building, it would be desirable to have a non-vanishing superpotential generated by worldsheet instantons in order to stabilize the bundle moduli in the low energy effective theory.
An algebraic procedure for computing $\text{Pfaff} ({\bar \partial}_{V_{\G_j} (-1)}) $ has been developed in~\cite{Buchbinder:2002ic, Buchbinder:2002pr}.
A non-zero Pfaffian implies \cite{Witten:1999eg} that the full contribution $W(\G_j)$ of a single curve is non-vanishing.
However, the determinant factors in the denominator depend on Calabi--Yau metric, which makes their computation impossible without differential geometry methods.
Given that these are still largely unknown for Calabi--Yau manifolds, we cannot in general rule out that different non-zero contributions do not cancel each other in the sum \eqref{1.6}.
In fact, for a broad class of models, a ``residue theorem'' \cite{Beasley:2003fx} by Beasley and Witten proves such a cancellation for all homology classes.

Therefore, the simplest examples in which we can make definitive statements about the (non-)vanishing of the superpotential are on Calabi--Yau geometries which have homology classes with a single isolated genus-zero curve---examples which we have constructed in the previous section.
If we can show that the contributions of such a curve is non-zero, then there cannot be any cancellations.\footnote{As mentioned previously, the vectors $\alpha_I ([\G])$ are linearly independent for linearly independent homology classes $[\G]$.
Hence the exponential prefactors $\exp(i \, \alpha_I([\G]) \, T^I)$ are in general independent functions of the K\"ahler moduli $T^I$ for independent homology classes and cannot cancel even if the Pfaffians are linearly dependent.}
Since the (non-)vanishing of $W(\G)$ for a single curve $\G$ is equivalent to the (non-)vanishing of $\text{Pfaff} ({\bar \pt}_{V_{\G} (-1)})$, we will now discuss how to compute it for spectral cover bundles $V$ on threefolds that are elliptically fibered over del Pezzo surfaces.


\subsection{A general method for computing the Pfaffian}


As was explained in~\cite{Witten:1999eg, Buchbinder:2002ic, Buchbinder:2002pr}, 
the zero modes of $\text{Pfaff} ({\bar \pt}_{V_{\G} (-1)})$ are global sections of $V_{\G} (-1)$, that is elements in 
$H^0 (\G, V_{\G} (-1))$.
The existence of such sections depends on the complex structure and, in particular, on the bundle moduli of $V$.
Therefore, as a function on the bundle moduli space ${\cal M}(V)$, $\text{Pfaff} ({\bar \pt}_{V_{\G} (-1)}) =0$ if and only if the cohomology group $H^0 (\G, V_{\G} (-1)) \neq 0$. 
Hence, our goal is to understand the dependence of its dimension $h^0 (\G, V_{\G} (-1))$ on the bundle parameters.

Recall from last section that we are working on an elliptically fibered Calabi--Yau $\pi: X \rightarrow B = dP_r$, with a vector bundle $V$ specified by spectral data $({\cal C, N})$.
We are interested in the case when the curve $\G$ is the lift $\cE$
of one of the exceptional curves $E$ in $dP_r$ by the zero section $\s$ of $X$. 
Just like in the previous section, let us denote by
$\Sigma= \pi^{-1} E$ the lift of the curve $E$ to $X$ and $\cE = \s \cdot \Sigma$ the curve of intersection with the zero section. 
We also denote by $\cC_{\S}= \cC|_{\S}$ the restriction of the spectral cover to $\S$ and $\cN_{\S}=  \cN|_{\S}$ the restriction of the line 
bundle $\cN$ to $\S$. 
Note that $\cC_{\S}$ is a spectral cover of the elliptically fibered surface $\S$ with covering map $\pi_{\cC_{\S}} = \pi|_\S: \cC_{\S} \to E$. It was shown in~\cite{Buchbinder:2002pr} that the vector bundle $V$ restricted to the curve $\cE$ 
(or any other curve sitting in the zero section $\s$) is given by 
\begin{align} 
V|_{\cE}= \pi_{\cC_{\S} *}  (\cN_{\S}|_{\cC_{\S}} )\,. 
\label{3.1}
\end{align}
Using a Leray spectral sequence we then obtain
\begin{align} 
h^0 (\cE, V|_{\cE} (-1))= h^0 (\cE, \pi_{\cC_{\S} *}  (\cN_{\S}|_{\cC_{\S}} ) \otimes \cO_{\cE} (-1)) =
h^0 (\cC_{\S}, \cN_{\S} (-F_{\S})|_{\cC_{\S}})\,, 
\label{3.2}
\end{align}
where $F_{\S}$ is the elliptic fiber of $\S$ and $\cN_{\S} (-F_{\S}) \equiv \cN_{\S} \otimes \cO_{\S} (-F_{\S})$.
To compute the last term we consider the short exact sequence 
\begin{align} 
0 \to \cN_{\S} (-F_{\S} - \cC_{\S}) \stackrel{f_{\S}}{\to} \cN_{\S} (-F_{\S})  \stackrel{r}{\to}   \cN_{\S} (-F_{\S})|_{\cC_{\S}} \to 0\,. 
\label{3.3}
\end{align}
Here $r$ is the restriction map and $f_{\S}$ is multiplication by a section which vanishes precisely on $\cC_{\S}$. 
Such a section is given by $f_{\S}= f|_{\S}$, where $f$ is the spectral cover equation~\eqref{2.3.5}. Note that $f_{\S}$ depends on the parameters 
of the spectral cover restricted to $\S$, that is on the moduli of the vector bundle $V$. 
The computation simplifies when the bundle satisfies
\begin{align} 
H^0 (\S, \cN_{\S} (-F_{\S} - \cC_{\S}) )= H^0 (\S, \cN_{\S} (-F_{\S}) ) =0\, ,
\label{3.4}
\end{align}
which will be the case in the explicit examples presented below.
Then the cohomology sequence of~\eqref{3.3} yields
\begin{align} 
0 \to H^0 (\cC_{\S}, \cN_{\S} (-F_{\S})|_{\cC_{\S}}) \to H^1 (\S, \cN_{\S} (-F_{\S} - \cC_{\S}) ) \stackrel{f_{\S}}{\to} H^1 (\S, \cN_{\S} (-F_{\S}) ) \to \dots\,. 
\label{3.5}
\end{align}
Hence, $H^0 (\cC_{\S}, \cN_{\S} (-F_{\S})|_{\cC_{\S}}) $ arises as the kernel of the linear map $f_{\S}$ which can be represented by a matrix. In examples of interest 
$f_{\S}$ is a square matrix, i.e., $h^1 (\S, \cN_{\S} (-F_{\S} - \cC_{\S}) ) = h^1 (\S, \cN_{\S} (-F_{\S}) ) $.
Then $H^0 (\cC_{\S}, \cN_{\S} (-F_{\S})|_{\cC_{\S}}) \neq 0 $ if and only if 
\begin{align} 
{\det} f_{\S} =0\,. 
\label{3.6}
\end{align}
The determinant $\det f_\S$, like $f_\S$, depends on the parameters of spectral cover equation $f=0$.
Therefore it defines a holomorphic section of a line bundle on the moduli space ${\cal M} (V)$ of the vector bundle $V$.
By \eqref{3.2}, the vanishing locus of this section in ${\cal M} (V)$ is where the Dirac operator acquires a zero mode, and is hence the same as the Pfaffian divisor, i.e., the divisor in ${\cal M}(V)$ where $\text{Pfaff} ({\bar \pt}_{V_{\cE} (-1)}) = 0$:
\begin{align}
	\{ \text{Pfaff} ({\bar \pt}_{V_{\cE} (-1)}) = 0 \} = \{\det f_\S = 0\} \subset {\cal M}(V)
\end{align}
This means that $\text{Pfaff} ({\bar \pt}_{V_{\cE} (-1)})$ is a section of the same line bundle as $\det f_\S$ with same vanishing locus.
Therefore, they have to be the same section up to a constant factor:
\begin{align} 
\text{Pfaff} ({\bar \pt}_{V_{\cE} (-1)}) \sim \det f_{\S} \, .
\label{3.7}
\end{align}
The proportionality constant depends on the Calabi--Yau and bundle metrics and hence cannot be computed in our algebraic approach. 
However, if $\det f_{\S} $ is not identically zero we conclude that $ \text{Pfaff} ({\bar \pt}_{V_{\cE} (-1)}) $ is non-zero.
Since $\cE$ is the only isolated genus-zero curve in its homology class, it also implies that the superpotential \eqref{3.5} is non-zero.


\subsection{Examples with non-zero Pfaffians}


\subsubsection[\texorpdfstring{$dP_1$}{dP1}]{$dP_1$}


Let us choose the base of $X$ to be the first del Pezzo surface $dP_1$. Examples with $B= dP_1= {\mathbb F}_1$ were extensively studied 
in~\cite{Buchbinder:2002ji, Buchbinder:2002ic, Buchbinder:2002pr} and we refer to these paper for additional details. 

The Mori cone of $dP_1$ is spanned by the two curves $E$ and $\ell -E$, where, to recall, $E$ is the blown up ${\mathbb P}^1$ and 
$\ell$ is the hyperplane divisor.\footnote{In~\cite{Buchbinder:2002ji, Buchbinder:2002ic, Buchbinder:2002pr} $E$ was denoted by $\cS$ and $\ell -E$ was denoted by $\cE$.}
Their intersection numbers are~\eqref{2.1}:
\begin{align} 
E^2=-1\,, \qquad (\ell -E)^2=0\,, \qquad E \cdot (\ell -E)=1\,. 
\label{4.1}
\end{align}
The only isolated curve in $dP_1$ is $E$, so the isolated curve in $X$ of interest is $\cE= \s \cdot \pi^{-1} E= \s \cdot \S$. 
We will choose the rank of the vector bundle to be three, which means that the spectral cover is of the form 
\begin{align} 
\cC= 3 \s + \pi^{-1} \eta\,, \qquad \eta = a E + b (\ell -E)\,. 
\label{4.2}
\end{align}
The spectral cover is effective if and only if $a, b \geq 0$. As was discussed previously, $\cC$ is an irreducible surface if 
$\eta \cdot E$ is non-negative and $\eta - 3 \overline{K}_{dP1}$ is effective. Using eqs.~\eqref{2.1}, \eqref{2.2.1} we find that it is equivalent to the following conditions 
\begin{align} 
b \geq a\,, \qquad a \geq 6\,, \qquad b \geq 9\,. 
\label{4.3}
\end{align}
Finally, to find positive spectral covers we demand that $\cC$ intersects positively all effective curves in $X$. A basis of effective curves in $X$
is given by $F$, $\s \cdot \pi^{-1}E$, $\s \cdot \pi^{-1}(\ell -E)$, so that it is enough to demand that $\cC$ intersects these three curves positively. 
Since $\cC$ is a triple cover of the base it intersects the elliptic fiber in three points, $\cC \cdot F=3$. The intersection of $\cC$ with 
the remaining two curves can be computed using the formula $\s \cdot \s =-\s \cdot \pi^{-1}\overline{K}_{dP1}$ for the zero section $\sigma$ of a Weierstrass model. 
This yields the following conditions:
\begin{align}
b -a >3\,, \qquad a > 6\,. 
\label{4.4}
\end{align}

Now let us specify the line bundle $\cN$ on $X$. For this we will choose the parameters $\l$ to be $\frac{3}{2}$. Then we obtain 
\begin{align}
\cC_{\S}= 3 \s_{\S} +  (b -a) F_{\S} \,, \qquad
\cN_{\S}= \cO_{\S} (6 \s_{\S}+ (5 - b +a) F_{\S} )\,,
\label{4.5}
\end{align}
where $\sigma_\Sigma = \sigma|_\Sigma$ for $\S = \pi^{-1} E$.
Let us now choose $b-a=5$ which is consistent 
with the spectral cover being effective, irreducible and positive. Then we get 
\begin{align}
\begin{split}
\cC_{\S} &= 3 \s_{\S} +  5 F_{\S}\,, \\
\cN_{\S} &= \cO_{\S} (6 \s_{\S})\,, \quad \cN_{\S}(-F_{\S}) = \cO_{\S} (6 \s_{\S} -F_{\S} )\,, \quad 
\cN_{\S}(-F_{\S}- \cC_{\S} ) =  \cO_{\S} (3 \s_{\S} - 6F_{\S} )\,.
\label{4.6}
\end{split}
\end{align}
This particular case was one of the examples studied in detail in~\cite{Buchbinder:2002ic, Buchbinder:2002pr}. We will not repeat the calculations 
since they are rather long and technical. 
Instead, we simply state the results, which show that the spectral cover $f_{\S}$ as well as the 
vector spaces $H^1 (\S, \cN_{\S} (-F_{\S} - \cC_{\S}) ) $ and $H^1 (\S, \cN_{\S} (-F_{\S}) )$ can be parametrized
in terms of homogeneous polynomials on the curve $E ={\mathbb P}^1$. The Pfaffian is non-zero and is given by a degree 20 polynomial 
\begin{align}
\text{Pfaff} ({\bar \pt}_{V_{\cE} (-1)}) \sim {\cal P}^4\,, 
\label{4.7}
\end{align}
where
\begin{align}
\begin{split}
{\cal P} = \, & \chi_1^2 \chi_3 \phi_3^2 - \chi_1^2 \chi_2 \phi_3 \phi_4 -2 \chi_1 \chi_3^2 \phi_3 \phi_1 
 \\
 - \, & \chi_1 \chi_2 \chi_3 \phi_3 \phi_2 +  \chi_2^2 \chi_3 \phi_1 \phi_3 +  \chi_1^3 \phi^2_4
 \\
 - \, & 2  \chi_3 \chi_1^2 \phi_2 \phi_4 +\chi_1  \chi_3^2  \phi_2^2   + 3 \chi_1 \chi_2 \chi_3 \phi_1 \phi_4
 \\
 + \, & \chi_1 \chi_2^2  \phi_2 \phi_4 + \chi_3^3 \phi_1^2 -   \chi_2 \chi_3^2  \phi_1 \phi_2 -\chi_2^3 \phi_4 \phi_1\,. 
\end{split}
\label{4.8}
\end{align}
The variables $\chi_i$, $\phi_j$ are parameters of the spectral cover, restricted to $E$. We refer to~\cite{Buchbinder:2002ic, Buchbinder:2002pr} for details and other examples.


\subsubsection[\texorpdfstring{$dP_2$}{dP2}]{$dP_2$}


Let us now choose the base $B$ to be $dP_2$. The Mori cone is spanned by the three curves $E_1$, $E_2$, $\ell-E_1-E_2$. 
Note that all three curves have self-intersection $-1$ and intersect $\overline{K}_{dP2} \equiv \overline{K}_2$ at one point.
We can use any of them to construct an isolated curve $\cE =\s \cdot \pi^{-1} E$ which is unique
in its homology class on $X$. For concreteness, let us choose $E= E_1$. As before, let us choose the rank of the vector bundle to be three, so that the
spectral cover is given by 
\begin{align} 
\cC= 3 \s + \pi^{-1} \eta\,, \qquad \eta = a E_1 + b E_2 + c (\ell -E_1 - E_2 )\,. 
\label{5.1}
\end{align}
The spectral cover is effective if and only if $a, b, c \geq 0$. To find irreducible spectral covers we demand that 
$ \eta - 3 \, \overline{K}_2$ is effective as well as 
\begin{align}
\eta \cdot E_1 \geq 0\,, \quad \eta \cdot E_2 \geq 0\,, \quad \eta \cdot (\ell -E_1- E_2) \geq 0\,. 
\label{5.2}
\end{align}
Using~\eqref{2.1}, \eqref{2.2.1} we obtain the following conditions for $\cC$ to be irreducible 
\begin{align}
\begin{split}
& a \geq 6\,, \qquad b \geq 6\,, \qquad c \geq 9\,,  \\
& c \geq a\,, \qquad c \geq b\,, \qquad a+ b \geq c\,. 
\end{split}
\label{5.3}
\end{align}
To find positive spectral covers we demand that $\cC$ intersects positively all effective curves in $X$. For this it is enough to consider
a basis of effective curves $F$, $\s \cdot \pi^{-1} E_1$, $\s \cdot \pi^{-1} E_2$, $\s \cdot \pi^{-1} (\ell -E_1- E_2)$, yielding the following conditions 
\begin{align}
c-a >3\,, \qquad c- b >3\,, \qquad a+ b- c >3\,. 
\label{5.4}
\end{align}
To describe the line bundle $\cN$ we again make the choice $\l=\frac{3}{2}$. Restricting $\cC$ and $\cN$ to $\S= \pi^{-1} E_1$ we obtain 
\begin{align}
\cC_{\S}= 3 \s_{\S} + (c-a)F_{\S}\,, \qquad \cN_{\S}= \cO_{\S}( 6 \s_{\S} + (5- c+a)F_{\S})\,. 
\label{5.5}
\end{align}
If we choose $c-a=5$ which is consistent with the spectral cover being effective, irreducible and positive, 
we find that $\cC_{\S}$ and $\cN_{\S}$ are the same as in eq.~\eqref{4.6}. 
Since the sequence~\eqref{3.5} depends only on $\cC_{\S}$, $\cN_{\S}$ and not on any additional data of $X$ or $V$ we conclude that
the calculation 
of the Pfaffian for this case is the same as in the previous subsection yielding the same non-vanishing result~\eqref{4.7}, \eqref{4.8}. 


\subsubsection[\texorpdfstring{$dP_3$}{dP3}]{$dP_3$}


Now let us choose the base $B= dP_3$. The Mori cone of $dP_3$ is spanned by the six curves
\begin{align}
C_k \in \{E_1, \ E_2, \ E_3, \ \ell - E_1- E_2,  \ \ell -E_2 - E_3, \ \ell - E_1- E_3\} \, ,
\label{6.1}
\end{align}
which have have self-intersection number $-1$ and intersect $\overline{K}_{dP_3} \equiv \overline{K}_3$ at one point. 
We can use any of them to construct an isolated curve $\cE =\s \cdot \pi^{-1} E$ which is unique
in its homology class.
As before, let us choose $E = E_1$ and the rank of the vector bundle to be three, so that the
spectral cover is given by 
\begin{align}
\begin{split}
\cC &= 3 \s + \pi^{-1} \eta\,, \\ 
\eta & = n_1 E_1 + n_2 (\ell - E_1- E_2) + n_3 E_2 + n_4 (\ell -E_1 - E_3 )+ 
n_5 (\ell - E_2 -E_3) 
+ n_6 E_3\,. 
\end{split}
\label{6.2}
\end{align}
The spectral cover is effective if and only if all $n_k \geq 0$. To find irreducible spectral covers we demand that 
$ \eta - 3 \overline{K}_3$ is effective as well as 
\begin{align}
\eta \cdot C_k \geq 0\,,
\label{6.3}
\end{align}
where the curves $C_k$ are given in~\eqref{6.1}. Explicit calculations give the following conditions
\begin{align}
\begin{split}
& n_k \geq 3\,, \qquad k=1, \dots, 6\,, 
 \\
& n_2 +n_4 \geq n_1\,, \qquad  n_1 +n_3 \geq n_2\,, \qquad n_2 +n_5 \geq n_3\,,  \\
& n_1 +n_6 \geq n_4\,, \qquad  n_3 +n_6 \geq n_5\,, \qquad n_4 +n_5 \geq n_6\,.
\end{split}
\label{6.4}
\end{align}
The positivity conditions in this case are
\begin{align}
\begin{split}
&
n_2 +n_4 > n_1+3 \,, \qquad  n_1 +n_3 > n_2+3 \,, \qquad n_2 +n_5 > n_3+3 \,,  \\
&
n_1 +n_6 > n_4+3 \,, \qquad  n_3 +n_6 > n_5+3 \,, \qquad n_4 +n_5 > n_6+3 \,.
\end{split}
\label{6.5}
\end{align}
With $\l=\frac{3}{2}$, we have
\begin{align}
\cC_{\S}= 3 \s_{\S} + (n_2+ n_4-n_1)F_{\S}\,, \qquad \cN_{\S}= \cO_{\S}( 6 \s_{\S} + (5- n_2- n_4 +n_1)F_{\S})\,. 
\label{6.6}
\end{align}
If we choose $n_2+ n_4-n_1 =5$ which is consistent with the spectral cover being effective, irreducible and positive, 
we find that $\cC_{\S}$ and $\cN_{\S}$ are the same as in eq.~\eqref{4.6}. Hence, this case yields a non-vanishing Pfaffian
given by the expression identical to~\eqref{4.7}, \eqref{4.8}.

Clearly, many such examples with a non-vanishing Pfaffian and, hence, a non-zero superpotential, can be found by appropriately choosing
the spectral cover and the line bundle, and performing calculations along the lines of~\cite{Buchbinder:2002ic, Buchbinder:2002pr}.
Unfortunately, even a slight change of the integers for the curve $\eta$, of the rank of the vector bundle, or of the parameter $\l$ 
in the line bundle $\cN$ quickly makes the number of spectral cover parameters and the size of the matrix $f_{\S}$ very large. 
In general, the Pfaffian in these models is a high degree polynomial depending on a large number of parameters.

The same analysis as above can also be performed to study vector bundles on other del Pezzo surfaces with $r>3$. 
However, the Mori cone for $r>3$ is generated by quite a large number of curves (see e.g.~\cite{Donagi:2004ia} for details) which makes the conditions for the spectral 
cover to be irreducible and positive combinatorially cumbersome.   
Furthermore, del Pezzo surfaces for $r\leq 3$ are always toric.
As we will see momentarily, this actually implies that in all our examples, the Calabi--Yau threefold $X$ can be embedded ``favorably'' into an ambient toric variety.
In this geometric setting, Beasley and Witten have shown \cite{Beasley:2003fx} that for a broad class of heterotic compactifications, the superpotential generated by a homology class of curves actually vanishes---which is clearly not the case in our examples.
In the following, we will resolve this apparent conflict by arguing that it is not the geometry but, rather, our vector bundles that do not fall into the category they have considered.


\section{Evading the Beasley--Witten residue theorem}\label{sec:beasley_witten}

In~\cite{Beasley:2003fx} (also see the earlier papers~\cite{Distler:1986wm, Distler:1987ee, Silverstein:1995re, Basu:2003bq}) 
Beasley and Witten showed that under certain assumptions the collective superpotential contribution \eqref{1.6} of all curves within the same homology class vanishes. 
Formulated as a residue theorem, it heavily restricts phenomenologically viable model building with such models.
Since our examples do not have vanishing superpotential, they cannot satisfy all of the assumptions required by the theorem.
In the following, we shall see that it is not the geometry---that is, the Calabi--Yau threefold $X$---but, rather, the choice of vector bundle that allows the Beasley--Witten theorem to be violated.

\subsection{The residue theorem}\label{sec:residue_review}


To begin with, let us briefly review the assumptions that go into the proof of the residue theorem.
Let $X \stackrel{i}{\hookrightarrow} \cal A$ be a complete intersection Calabi--Yau 
manifold in the ambient space ${\cal A}$ which is toric.
Additionally, we assume that the stable vector bundle $V$ on $X$ 
is obtained as a restriction of a stable vector bundle $W$ on ${\cal A}$, $V=W|_X$.
It was shown by Beasley and Witten that if these assumptions are 
satisfied the sum~\eqref{1.6} of all superpotential contributions from isolated rational curves vanishes for any homology class. 
This result was interpreted in~\cite{Beasley:2003fx} as a residue theorem.
By formulating the worldsheet instanton as a ``half-linear'' sigma model, they could show that the superpotential contributions of individual curves within the same homology class can be interpreted as residues of a meromorphic function on the moduli space.
Within their assumptions, the moduli space is compact, and hence the sum of residues must vanish.

The description through a sigma model makes the necessity of a toric ambient space ${\cal A} \supset X$ with a stable parent vector bundle apparent.
As was pointed out in \cite{Buchbinder:2016rmw}, however, the analysis of Beasley and Witten actually implicitly assumes 
that the K\"ahler form $\o_X$ on $X$ is obtained as a restriction, $\o_X =\o_{{\cal A}}|_X$, of the ambient space K\"ahler form $\o_{{\cal A}}$.
If the Calabi--Yau manifold is favorable \cite{Anderson:2008uw}, that is, the entire 
second cohomology of $X$ descends from the ambient space---or, equivalently, $H^{1,1}(X,\bbR) = i^* (H^{1,1}({\cal A}, \bbR))$---then this assumption is indeed satisfied for all choices of K\"ahler form $\o_X$.
In this setting, the cancellation of instanton contributions holds for curves that are homologous on $\cal A$.

The first attempts to evade the residue theorem were carried out within the context of a Schoen threefold $\tilde{X}$ very similar to, but not quite identical with, that of the ``heterotic standard model'' introduced in \cite{Braun:2005bw,Braun:2005zv,Braun:2005nv}. 
This threefold is a CICY inside ${\cal A} = \bbP^2 \times \bbP^2 \times \bbP^1$ with $h^{1,1}({\cal A}) = 3$.
Since the Schoen famously has $h^{1,1}(\tilde{X}) = 19$, this embedding cannot be favorable.\footnote{In fact, it was shown in \cite{Anderson:2017aux} that the Schoen cannot be embedded favorably in any product of projective spaces.}
Because of that, it was shown \cite{Buchbinder:2016rmw} that indeed, after introducing a vector bundle, the superpotential is non-zero in this case.
This observation was strengthened by the results of \cite{Buchbinder:2017azb}, where several completely different heterotic vacua were presented with non-vanishing instanton superpotentials, all due to an unfavorable embedding.

A more subtle way of obtaining a non-zero superpotential is to consider threefolds $X$ with non-trivial torsion in $H_{2}(X,\mathbb{Z})$.
As discussed in Section \ref{sec:general_structure_superpotential}, this can lead to a different torsion factor $\chi$ for curves in the same real homology class $[\G] \in H_2(X,\bbR)$ which can spoil the cancellation among contributions within $[\G]$.
Such a threefold has been constructed in \cite{Braun:2007tp, Braun:2007xh, Braun:2007vy}, which was obtained as a $\bbZ_3 \times \bbZ_3$ quotient of the same Schoen manifold $\tilde{X}$ discussed above.
Due to the quotient, one has $\text{Tors}(H_{2}(X,\mathbb{Z})) = \bbZ_3 \times \bbZ_3$.
By solving for the Gromov-Witten invariants of this quotient threefold, it was shown that a subset of these torsion classes each contained only a single isolated curve.
In the presence of a vector bundle, it was shown \cite{Buchbinder:2016rmw} that each such curve contributed a non-zero instanton which, furthermore, did not cancel when summed together.
The appearance of torsion in the quotient actually can be traced back to the non-favorability of the covering Schoen threefold $\tilde{X}$.
Through the quotient, the ``non-favorable'' curves classes of $\tilde{X}$---that is, curves Poincar\'{e}-dual to those K\"ahler forms that did not come from the ambient space---manifest themselves through $\text{Tors} (H_2) \neq 0$.

Given that our examples indeed produce non-vanishing superpotentials, one might suspect that they fall into this class of non-favorable models.
However, as we will now show, the Calabi--Yau threefolds we considered in the previous section can be embedded favorably into a toric ambient space.

\subsection{Favorable embedding of the Weierstrass model}

Our geometric setup---$X$ being a generic elliptic fibration over a toric del Pezzo surface $dP_r$, $r\leq 3$---satisfies the assumptions of the residue theorem, namely that $X$ has a favorable toric embedding.
In fact, such an embedding is naturally provided by the Weierstrass equation \eqref{eq:weierstrass_equation}.
This equation defines a hypersurface in the projective bundle
\begin{align}
	{\cal A} = \bbP ( \underbrace{\overline{K}_B^{\otimes 2}}_{\sim x} \oplus  \underbrace{\overline{K}_B^{\otimes 3}}_{\sim y} \oplus \underbrace{\cO_B}_{\sim z}) \, \stackrel{\phi}{\longrightarrow} B \, .
\end{align}
Because of the trivialization property of fiber bundles, $\cal A$ is smooth if $B$ is.
Furthermore, if $B$ is toric, then $\cal A$ is also a toric space (see e.g.~\cite{cox2011toric}).
These conditions are satisfies for $B = dP_r$ with $r\leq 3$.
Smooth toric spaces have a particularly simple (co-)homology.
First, all (co-)homology groups of odd degree are zero.
For even degrees, we start with $H_{d-2}({\cal A},\bbR)$ with $d = \dim_\bbR \cal A$, which is generated by toric divisor.
All higher homology groups are then generated by intersection products of these divisors modulo homological equivalence.

For the case at hand, the toric ambient space $\cal A$ has $r+2$ independent divisors:
Because $\cal A$ is a fiber bundle over $B = dP_r$, the $r+1$ independent divisors of $B$ pull-back to independent divisors of $\cal A$.
In addition, there is the ``hyperplane'' class $S$ of the fiber $\bbP^2$; in terms of equations it is linearly equivalent to the vanishing locus of a linear polynomial in the fiber coordinates $[x:y:z]$.
Therefore, we have
\begin{align}\label{eq:divisors_toric_ambient}
\begin{split}
	H_6 ( {\cal A}, \bbR) & = (\bbR\,S) \oplus \phi^{-1}(H_2(B, \bbR)) \\
	\stackrel{\text{Poincar\'{e}-dual}}{\Longrightarrow} \qquad H^2({\cal A}, \bbR) & = (\bbR \, \text{PD}(S) ) \oplus \phi^*(H^2(B,\bbR)) \, .
\end{split}
\end{align}

To determine $H_4({\cal A}) \cong H^4(\cal A)$, we have to take the intersection product (cup product for cohomology) of \eqref{eq:divisors_toric_ambient}.
For that, the only additional input we need to know is that $S \cdot S$ is homologous to $S \cdot \phi^{-1}(D_B) \in S \cdot \phi^{-1}(H_2(B,\bbR))$ for some divisor $D_B$ in the base, whose specific form is irrelevant.
Keeping in mind that divisors on $B$ intersect each other at points, over which the $\bbP^2$-fiber of ${\cal A}$ defines a 4-cycle class $\Phi$, we have
\begin{align}\label{eq:four-cycles_toric_ambient}
	\begin{split}
		H_4 ({\cal A}, \bbR) = H_6 ({\cal A},\bbR) \cdot H_6 ({\cal A},\bbR) = S \cdot \phi^{-1}(H_2(B, \bbR)) \oplus (\bbR \, \Phi) \cong H^4({\cal A}, \bbR) \, .
	\end{split}
\end{align}

Turning to the elliptic fibration, recall that a generic Weierstrass model $X \stackrel{\pi}{\rightarrow} B$ inside this ambient space is smooth and has no reducible fibers.
Furthermore, it also has a trivial Mordell--Weil group.
As a consequence, the Shioda--Tate--Wazir theorem \cite{MR2041769} implies that the space of divisors of $X$ is spanned by the zero section $\sigma$ and vertical divisors, i.e., pull-backs from the base.
An important feature of the Weierstrass model,
\begin{align}\label{eq:weierstrass_equation_2}
	y^2 z = x^3 + g_2 x z^2 + g_3 z^3 \, ,
\end{align}
is that the zero section $\sigma$, marking the point $[x:y:z] = [0:1:0]$ on the elliptic fiber, can be viewed as one of the three points of intersection in the $\bbP^2$ fiber of $\cal A$ between the Weierstrass equation and a hyperplane, e.g.~$x=0$.
Therefore, the divisor class $S \in H_6({\cal A})$ restricts to $3\sigma \in H_4(X)$.
Since $\cal A$ and $X$ share the same base, the vertical divisors of $X$ are obviously restrictions of divisors on $\cal A$ as well.
We have thus just shown that all divisors, and by Poincar\'{e}-duality also all K\"ahler classes of $X$ arise as restrictions from ${\cal A}$.
In fact, their dimension agree, so $H_4(X,\bbR) \cong H_6({\cal A}, \bbR)$. Hence, $X \subset \cal A$ is favorable.

For the subsequent discussion, it is useful to explicitly determine the 2-cycles on $X$ as well.
Since $X$ is a threefold, 2-cycles are dual to divisors.
The independent classes are
\begin{align}
	H_2(X, \bbR) = \sigma \cdot \pi^{-1}(H_2(B, \bbR)) \oplus (\bbR \, F)\, ,
\end{align}
where $F$ is the class of the elliptic fiber.
Note that this is again isomorphic to the restriction of 4-cycle classes \eqref{eq:four-cycles_toric_ambient} on ${\cal A}$, since $S |_ X  = 3\sigma$ and $\Phi|_X = F$.

\subsection{Evading the residue theorem with irreducible spectral cover bundles}\label{sec:chern_class_argument}

Since the geometry satisfies the assumptions of the Beasley--Witten residue theorem, the only possible explanation why our examples have non-vanishing superpotentials is that the bundles we consider do not arise as a restriction of a stable vector bundle on a toric ambient space.
For the ``natural'' choice of ambient space in terms of the $\bbP^2$ bundle of the Weierstrass model, this can indeed be shown explicitly.

At first glance, one might be surprised that our spectral cover constructions are so restrictive.
After all, any stable bundle $V$ on an elliptic fibration $X$ can be Fourier--Mukai-transformed into spectral data $({\cal C , N})$.
The crucial property of our examples is that ${\cal C}$ is irreducible and smooth.
With this assumption, the second Chern class takes the schematic form (see \eqref{2.8})
\begin{align}\label{eq:schematic_form_c2(V)}
	c_2(V) = \sigma \cdot \pi^{-1}(D_1) + \pi^{-1}(D_2 \cdot D_3) = \sigma \cdot \pi^{-1}(D_1) + k\,F
\end{align}
for some divisors $D_i \in H_2(B)$, and $k = D_2 \cdot D_3$ as intersection number in $B$.

Let us assume that there is a stable rank $n$ vector bundle $W$ on ${\cal A}$ that restricts to $V$ on $X$.
Then, we necessarily have to have $c_2(W) |_X = c_2(V)$.
However, as we have elaborated above, $H_4({\cal A})$, in which (the Poincar\'{e}-dual of) $c_2(W)$ lives, is isomorphic to $H_2(X) \ni \text{PD}(c_2(V))$.
Hence, we must have
\begin{align}\label{eq:c2_W}
	c_2(W) = \frac{1}{3}\,S \cdot \phi^{-1}(D_1) + k \, \Phi \, .
\end{align}
Now consider the restriction of $W$ to the fiber $\Phi$ of ${\cal A}$, which gives rise to a stable bundle on $\Phi \cong \bbP^2$.
In general, the second Chern class of a bundle on $\bbP^2$ is simply (Poincar\'{e}-dual to) a collection of points.
In our case, the number of these points is
\begin{align}
	c_2(W|_\Phi) = \Phi \cdot c_2(W) = \phi^{-1}({\text{pt}}) \cdot ( \Sigma \cdot \phi^{-1} (\tilde{D}_1) + k \, \phi^{-1}(\text{pt})) = 0 \, .
\end{align}
Furthermore, because we have $SU(n)$ bundles, $c_1(V)=0$, which because of the particular embedding also implies
\begin{align}
	c_1(V) = c_1(W) = 0 \quad \Longrightarrow \quad c_1(W|_\Phi) =0 \, .
\end{align}

By the Donaldson--Uhlenbeck--Yau theorem, a stable rank $n$ bundle on $\bbP^2$ with trivial first and second Chern classes has to be a sum of $n$ trivial line bundles.
However, in our setup, this would imply that the restriction of the bundle $V$ to the elliptic fiber of $X$ is also trivial:
\begin{align}
	V|_F = (W|_X)|_F = (\underbrace{{W|_\Phi}}_{\text{trivial}} )|_F \, .
\end{align}
However, the spectral cover bundle is by construction non-trivial on the elliptic fiber $F$:
It restricts to a sum of $n$ non-trivial, degree-zero line bundles $F$, which is specified by the $n$ points on $F$ marked by the $n$-sheeted cover $\cal C$.
We therefore arrive at the conclusion that there cannot be a stable vector bundle $W$ on ${\cal A}$ which restricts $V$.

Thus, the spectral cover models we presented in the previous section do not satisfy the Beasley--Witten assumptions with respect to the favorable toric embedding $X \subset \cal A$.
This is consistent with our result that the Pfaffians on $\Gamma$ and, therefore---because $\Gamma$ was unique in its homology class---the superpotential, is non-zero.

Note that our argument for the bundle not extending to the ambient space $\cal A$ is limited to this particular choice of toric embedding.
For example, if there is another favorable embedding $X \subset \cal A'$, where $h^{1,1}({\cal A'}) > h^{1,1}(X)$, then the Chern-classes of a putative bundle $W'$ on $\cal A'$ which restricts to $V$ can have terms which restrict trivially on $X$.
Thus, just knowing the Chern-classes of $V$ on $X$ would not allow us to make any predictions about the structure of $W'$.
In such a setting, one must find other arguments to disprove the existence of $W'$.
In fact, given the validity of the residue theorem, one could view the superpotential itself as an obstruction of $V$ to extend to a bundle $W$ on any favorable toric embedding.
In the case of threefolds $X$ where curve classes have unique isolated representatives, this would establish an explicit connection between the Pfaffian and the cohomology classes that parametrize the obstruction.
This could open up a novel, algebraic approach to understanding---and also evading---the Beasley--Witten theorem.


\section{Summary and outlook}\label{sec:summary}

In this paper we have presented a class of toy models for 4D heterotic compactifications that have a non-zero superpotential for the vector bundle moduli.
The compactification spaces are Calabi--Yau threefolds that are elliptically fibered over del Pezzo surfaces $dP_r$.
By lifting the exceptional curves on $dP_r$ with the zero section of the fibration, we obtain a set of holomorphic, isolated genus-zero curves $\cE$ which are unique in their homology class. On these spaces, we then construct stable vector bundles $V$ using spectral data $(\cal C, N)$.
Employing the algebraic methods developed in \cite{Buchbinder:2002ic, Buchbinder:2002pr}, we compute the associated Pfaffian factors for string instantons on the curves $\cE$, showing explicitly that they do not vanish. 
Since these curves are the only isolated genus-zero representatives within their homology class, the full superpotential generated by that class has to be non-zero.

Naively, this seems to be in contradiction with the residue theorem of Beasley and Witten, which guarantees, under certain assumptions, the cancellation of all superpotential contributions from homologous curves.
Indeed, as we have shown, some of our Calabi--Yau manifolds do satisfy the assumptions that they can be favorably embedded into a toric ambient space.
However, the paradox is resolved because the vector bundles we considered do not fall into the category that Beasley and Witten considered; that is, they are not restrictions of vector bundles on the toric ambient space.

One caveat of our analysis is that we have restricted ourselves to a single embedding space, which for elliptic fibrations is always a natural choice.
A valid concern is if there could be other favorable toric embeddings that have a  vector bundle on the ambient space that does restrict to the bundle on the Calabi--Yau threefold.
Our approach does not straightforwardly extend to such cases, and one would have to analyze them one by one, perhaps using different arguments than those presented in Section \ref{sec:chern_class_argument}.
From another perspective however, the scenarios analyzed here might shed light on an algebraic interpretation of the residue theorem, at least on geometries with isolated rational curves that are unique in their homology class.
Our findings suggests that the non-vanishing of the superpotential could be understood as an obstruction for the vector bundle $V$ to extend to \textit{any} favorable toric embedding space.
It would be interesting to quantify these obstructions and relate them to the Pfaffian, thereby explicitly arguing against $V$ extending to other possible toric embeddings.
Another exciting direction along the same lines is to analyze the superpotential in the dual F-/M-theory setting \cite{Donagi:1996yf, Curio:1997rn, Anderson:2015yzz , Cvetic:2012ts, Braun:2018fdp}.
In particular, recent advances \cite{Candelas:1996su, Andreas:1998zf, Berglund:1998ej, Anderson:2014gla, Cvetic:2015uwu, Cvetic:2016ner} on heterotic/F-theory duality uses toric embeddings of the compactification spaces, which is therefore naturally tied to the Beasley--Witten theorem.
Since in the dual F-theory the bundle data is also geometrized, one might hope that this could provide a more direct algebro-geometric approach to the residue theorem.


\section*{Acknowledgements}
The authors are indebted to Antonella Grassi, Andre Lukas, Tony Pantev and Fabian Ruehle for valuable discussions. 
The work of Ling Lin and Burt Ovrut is supported in part by the DOE under contract No.~DE SC0007901.


\newpage
\bibliography{het_lit.bib}{}
\bibliographystyle{JHEP}

\end{document}